\documentstyle[twocolumn,prl,aps,epsfig]{revtex}

\begin{document}
\draft
\twocolumn[\hsize\textwidth\columnwidth\hsize\csname  
@twocolumnfalse\endcsname                             

\title{Fabrication of mesoscopic Nb wires using conventional 
e-beam lithographic techniques}
\author{Nam Kim, Klavs Hansen, Jussi Toppari, 
Tarmo Suppula, and Jukka Pekola}
\address{Department of Physics,
University of Jyv\"{a}skyl\"{a} 
P.O.Box 35, FIN-40351 Jyv\"{a}skyl\"{a}, Finland }
\maketitle

\begin{abstract}
Conventional electron beam lithography has been used 
to fabricate mesoscopic Nb wires with a superconducting transition 
temperature above 7.0 K. The typical line width and the thickness were 200 nm 
and 45 nm respectively. Nb was deposited 
in an ultra-high vacuum evaporation chamber using electron gun heating. 
All samples exhibited a normal-superconducting transition.
The transition temperature decreased with thickness 
and line width. 
To demonstrate the feasibility of two angle evaporation techniques we also
fabricated small Nb/(Al-)AlO$_{x}$/Nb 
tunnel junctions.
\end{abstract}

 
] 
\narrowtext 

A number of 
nanometer-scale devices such as single-electron transistors \cite{devoret}, 
superconducting quantum bits \cite{nakamura}, 
and other mesoscopic superconducting devices \cite{hekking} have been 
realized using the self-alignment technique which provides submicron accuracy \cite{dolan}. 
Self alignment 
is achieved by the shadow evaporation technique, commonly used with the 
resist polymethylmethacrylate (PMMA) and co-polymer [P(MMA-MAA)] 
as a double layer stencil mask patterned by electron-beam (e-beam) 
lithography. Until now this conventional shadow evaporation 
technique has been applied successfully for the soft metals 
such as Al, Cu and Pb. For the refractory metals 
such as Nb, W or Ta, however, this technique 
is known to be difficult to apply. In particular Nb is a promising 
alternative to the soft metals for superconducting nano devices 
such as single-electron transistors and quantum bits,
because of its large 
superconducting gap and high stability under thermal cycling. 
A demonstration that it is possible to fabricate
nanostructures by conventional e-beam lithography would therefore
be highly significant.

The problems with the application of the 
conventional technique have been 
ascribed to the partial decomposition of the PMMA-co-polymer double 
layer during the evaporation of the refractory metals \cite{harada}. 
The resulting outgassing from the resist \cite{dubos} and consequent contamination 
of the deposited Nb would then explain the changes in the electronic 
properties of the deposited Nb. Contrary to this common experience,
we were able to fabricate mesoscopic Nb wires with zero-field 
critical temperatures $T_{c}$ higher than 7.0 K by using the conventional 
shadow evaporation technique.

The results presented here were obtained with an e-beam 
lithography process which did not have any special 
features but followed the most common procedures 
with conventional recipes. We deposited a double layer 
of  PMMA-P(MMA-MAA) on the oxidized Si substrate with 
the thickness of the oxidization layer of about 250 nm. 
The spinning rates for PMMA and P(MMA-MAA) were 3000 rpm 
and 6000 rpm, respectively, and the spinning time was 30 
seconds for both. The thickness of the PMMA and P(MMA-MAA) 
were measured to be about 270 nm and 300 nm, respectively. 
The resist was baked at $160 ^\circ$C for about 60 minutes. 
We then drew a pattern of wire (see Fig. 1) by using a 
scanning electron microscope (JEOL, JSM 840A). The 
line width was about 200 nm. To develop the upper layer 
of the PMMA resist we immersed the sample in a mixed 
(1:2) solution of methyl {\it iso}-butyl ketone and 
isopropylic alcohol for 12 seconds. To develop the 
lower layer of the P(MMA-MAA) resist and to form an undercut, 
the sample was dipped into a solution (1:2) of methyl glycol 
and methanol for 15 seconds.

The Nb (99.9 $\%$, Goodfellow) was evaporated onto the 
patterned substrate in an ultra-high vacuum (UHV) 
chamber equipped with a cryo-vacuum pump (Cryo-Torr High 
vacuum pump, CTI-Cryogenics) and a liquid nitrogen trap. 
The pressure of the UHV chamber was $2\times10^{-8}$ mbar,
the evaporation rate was about 0.45 nm/s and the power of the
electron gun was 2 kW. A crucial feature in the evaporation chamber 
was the 40 cm distance between the substrate 
and the Nb crucible. This relatively large separation reduced the
indirect heating of the samples when evaporating Nb. All samples 
were made using the same evaporation parameters.

We have measured the change of sample resistance with temperature, 
$R(T)$, for the five samples fabricated (see Fig. 2 and Table 1). 
The typical dimension of the wires is about 200 nm in width and 11 $\mu$m in 
length while the width of the 2D film is about 10 $\mu$m. 

\begin{figure}[hb]
\center
\epsfig{file=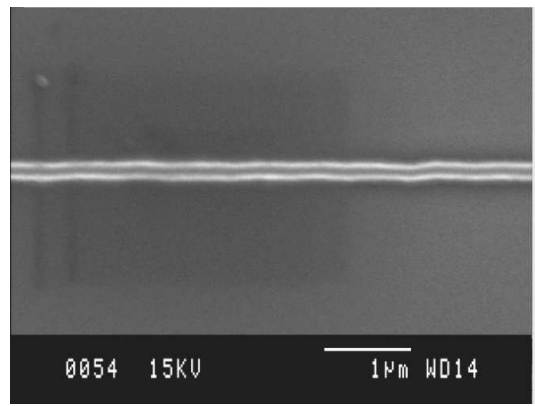,width=70truemm}
\vskip 5mm
\caption{Scanning electron micrograph of a Nb wire with 
thickness 45 nm and width 200 nm.}
\label{schem}
\end{figure}

\begin{figure}[h]
\center
\epsfig{file=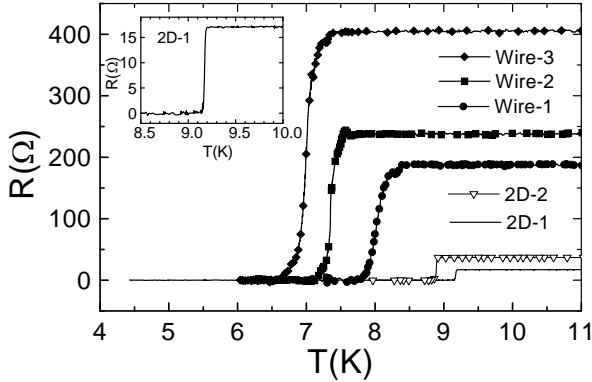, width=84truemm}
\vskip 3mm
\caption{Resistance vs temperature for various samples.
Inset: resistance vs temperature for 2D-1 on an expanded scale. }
\label{re}
\end{figure}

Details of sample parameters are summarized in Table 1. The sample 
resistance has been measured by the conventional 
lock-in amplifier technique in a four probe measurement configuration. 
The bias current level was 10 nA for wires and 100 nA for 2D films. 
A calibrated diode sensor (DT470, Lake Shore Cryotronics) 
was used to monitor the temperature.

Note that all the samples become superconducting, 
but with a transition temperature $T_{c}$ which depends 
on the sample dimension. One of the samples, the 2D Nb film with 
a thickness of 135 nm (2D-1) showed $T_{c}=9.1$ K, close 
to that of bulk Nb, 9.26 K (see the inset of Fig. 2). 
$T_{c}$ was defined as the temperature where the sample resistance 
is reduced to one half of its resistance at 10 K. We have found 
that the superconducting transition temperature decreases 
with Nb thickness or width as shown in Fig. 2. 
Wire-3 with dimensions typical of those in applications  
had a $T_{c}$ of 7.0 K and the resistivity ratio 
$\rho_{295 K}/\rho_{10 K}$ of 1.5.

\begin{figure}[t]
\center
\epsfig{file=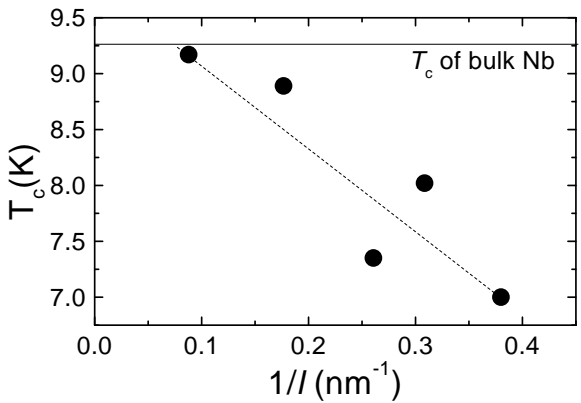,width=84truemm}
\vskip 2mm
\caption{$T_{c}$ vs inverse mean free path $1/l$ at 10 K. 
The mean free path $l$ at 10 K has been calculated 
from the formula $\rho l=8.7\times 10^{-12}$ $\Omega$cm$^{2}$. 
$T_{c}$ of bulk Nb is 9.26 K. The dashed line is a guide for the eye.}
\label{rg}
\end{figure}

\begin{figure}[ht]
\center
\epsfig{file=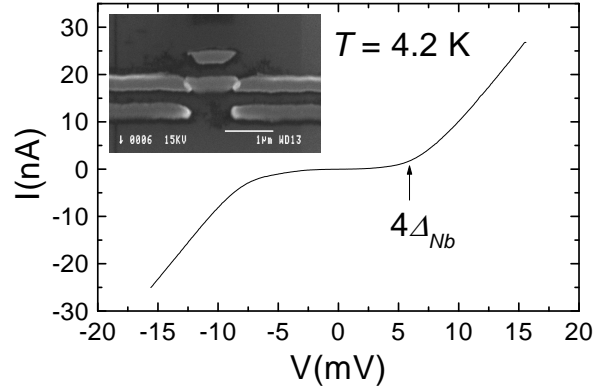,width=84truemm}
\vskip 1mm
\caption{$I-V$ characteristic of the Nb/(Al-)AlO$_{x}$/Nb junctions 
at 4.2 K. $\Delta_{Nb}=1.5$ meV. 
Inset: scanning electron micrograph of the Nb/(Al-)AlO$_{x}$/Nb 
junctions fabricated by the conventional two angle 
evaporation technique. The length of the white bar is 1 $\mu$m.}
\label{mittaus}
\end{figure}

The reason for the reduction of $T_c$ from the bulk value 
is worth considering. For this purpose we plot $T_{c}$ 
vs the inverse mean free path $1/l$ in Fig. 3 
where $l$ has been calculated from the resistivity at 10 K 
using the Drude formula $\rho l=8.7\times 10^{-12}$ 
$\Omega$cm$^{2}$ for Nb \cite{ashcroft}. The Fig. 3 shows 
that $T_{c}$ is decreased as the mean free path $l$ decreased. 
Our results are similar to the previous 
experimental observations \cite{crow} where the authors concluded 
that $T_{c}$ depends on the disorder of the sample. 
The authors of Ref. 8 suggested that 
disorder in Nb smears out the density of states of Nb 
metal and consequently reduce the density of states at 
the Fermi level, resulting in a decrease of  $T_{c}$. 
Since all our 
samples have been evaporated under an equal evaporation 
condition, the reduction of $T_{c}$ of our samples is 
consistent with the suggestion that it depends mainly 
on disorder of the sample.

To demonstrate the feasibility of the conventional two angle evaporation technique
and to take advantage of the high critical temperatures reached 
we have also fabricated Nb/(Al-)AlO$_{x}$/Nb junctions with a 
single electron transistor geometry (see the inset of Fig. 4). 
The 15 nm thick Al layer was evaporated on the 
67 nm thick Nb layer. 
The Al layer was subsequently oxidized in a static oxygen pressure of 
100 mbar for 5 minutes. Then the third layer of 
Nb was deposited at a different angle. 
We measured the current vs voltage ($I-V$) characteristic at 4.2 K
for the Nb/(Al-)AlO$_{x}$/Nb junctions  with a room temperature 
junction resistance 60 k$\Omega$ (Fig. 4). The $I-V$ curve 
clearly shows a superconducting gap, the size of which is very 
close to $4\Delta_{Nb}$ where $\Delta_{Nb}$ is the bulk Nb superconducting 
gap energy ($\Delta_{Nb}=1.5$ meV). These results are promising
for the fabrication of Nb-based nano-scale devices. 

In conclusion, using the conventional e-beam lithographic 
technique we have fabricated mesoscopic Nb wires with a 
width of about 200 nm showing $T_{c}$ higher than 7.0 K. 
In addition we could also make mesoscopic Nb/(Al-)AlO$_{x}$/Nb 
tunnel junctions showing a superconducting gap close to that 
of bulk Nb at 4.2 K.

This work has been supported by the Academy of Finland 
under the Finnish Center of Excellence Programme 2000-2005
(Project No. 44875, Nuclear and Condensed Matter Programme at
JYFL) and the EU (contract IST-1999-10673).
The authors thank Kurt Gloos and Jinhee Kim for discussions.

\begin{table}[b]
\caption{Results of the various samples. 
We denoted as 2D films the macroscopic samples 
with width 10 $\mu$m. $w$, $t$, $L$, $\rho$ and $l$ are the sample width, 
thickness, length,  resistivity and mean free path at 10 K. 
$\rho _{295 K}/\rho _{10 K}$ is the resistivity ratio for the two  
temperatures, 295 K and 10 K.}
\begin{tabular}{cccccccc}
Sample&$w$&$t$&$L$&$T_{c}$&
 $\rho$&$l$&${\rho _{295 K}}/{\rho _{10 K}}$\\ 
 &($\mu$m)&(nm)&($\mu$m)&(K)&
 ($\mu \Omega$cm)&(nm)&\\
\hline
2D-1&10&135&300&9.1&7.6&11&\\
 2D-2&10&67&160&8.9&15&5.7&2.2\\
 Wire-1&0.23&67&11&8.0&27&3.2&\\
 Wire-2&0.23&45&11&7.3&23&3.8&1.6\\
 Wire-3&0.20&45&11&7.0&33&2.6&1.5\\
\end{tabular}
\end{table}


\end{document}